%% file: main.tex
\documentclass[runningheads]{llncs}
\usepackage[T1]{fontenc}
\usepackage[utf8]{inputenc}
\usepackage{physics}
\usepackage{hyperref}
\usepackage{enumitem}
\usepackage{amsfonts}
\usepackage{orcidlink}
\usepackage{pythonhighlight}
\usepackage{fontawesome5}

\include{definitions.tex}

\usepackage{cleveref}
\usepackage{array}
\usepackage{graphicx}

\usepackage{float}

\newcommand{\kingdon}[0]{\pyth{kingdon}}
\newcommand{\ganja}[0]{\pyth{ganja.js}}

\begin{document}

\title{The Willing \textbf{Kingdon} Clifford Algebra Library}

\author{Martin Roelfs\inst{1}\orcidlink{0000-0002-8646-7693}\href{https://github.com/tBuLi/kingdon}{\faGithub}}
\institute{Flanders Make
\email{martin.roelfs@flandersmake.be}
\url{http://tbuli.github.io/teahouse}
}
\authorrunning{Martin Roelfs}

\maketitle

\begin{abstract}
Kingdon is an open-source Python package designed to seamlessly integrate Geometric Algebra (GA) into existing workflows. Unlike previous GA libraries, \kingdon{} is input-type-agnostic, and hence supports GA's over e.g. PyTorch tensors, NumPy arrays, or SymPy symbolic expressions, to name but a few. Despite this refusal to specialize, it delivers high performance by symbolically optimizing operators and leveraging input sparsity for Just-In-Time compiled expressions. Additionally, its visualization capabilities in Jupyter notebooks using \ganja{} align with the rapid prototyping workflow common to scientific research.
\end{abstract}


\section{Introduction}

The Python programming language is very popular in the scientific community - amongst other things - because of its ease of use and rapid prototyping.
This has resulted in a flourishing ecosystem of tooling for e.g. data analysis such as \verb|numpy|, \verb|scipy| and \verb|pandas|; machine learning such as \verb|PyTorch| and \verb|TensorFlow|, and algebra packages such as \verb|SymPy|.
Many of these tools are used to analyse problems involving geometry, in fields ranging from computer vision to astronomy; problems whose solutions would benefit from Geometric Algebra (GA) support.
However, existing GA projects in Python tend to be aimed towards one specific purpose. For example, the popular \verb|clifford| library \cite{python_clifford} assumes the input to be numerical, while the \verb|galgebra| library \cite{galgebra} can only be used for symbolic manipulations.
While these specialized packages are powerful within their own domain, they are not conducive to the rapid prototyping workflow for which Python is famous. 
To illustrate, it is not possible to formulate algorithms using simple numerical examples or symbolical inputs, and to then simply change the input to \verb|PyTorch| tensors containing large amounts of data without having to change the implementation of the algorithms themselves.

This highlights the main motivation behind the development of the \verb|kingdon| package: 
to enable programmers to write high level code that captures only the geometry, without additionally having to commit to a specific datatype or dependencies.
Therefore \verb|kingdon| was designed to be agnostic to input types, such that the exact same code will work on any datatype.
Additionally, \verb|kingdon| was written with performance in mind: it symbolically optimizes GA's unary and binary operators and leverages the sparseness of the input in order to create Just-In-Time compiled expressions of optimal computational complexity. 
While this on-the-fly code generation incurs some (typically small, see \cref{tab:codegen}) initial cost upon first execution, 
subsequent evaluations will be faster, 
and so when the goal is to e.g. train a neural network or analyse astronomically large amounts of data, 
the elimination of unnecessary operations will certainly pay off.
Furthermore, in keeping with the rapid prototyping mindset, \verb|kingdon| can visualize scenes within \verb|jupyter| notebooks using the popular \verb|ganja.js| package \cite{ganja}.

\section{Project Vision}

The \kingdon{} project aims to lower the threshold of GA adoption, both for new projects and existing code-bases. The project is released under the permissive MIT licence, and fully embraces the principles of open-source code. 
The ambition is to grow \kingdon{} into the de-facto Python GA package, and to grow into a community driven project.
Moreover, \kingdon{} is developed in conjunction with \texttt{ganja.js}'s successor \texttt{GAmphetamine.js}, to create a homogenous ecosystem across languages and platforms.

\section{Design Philosophy Principles}
    \begin{description}[style=unboxed,leftmargin=0cm]
        \item[Support GA's of arbitrary signature] \pyth{kingdon} supports GA's of arbitrary signature \GR{p,q,r}, where $p$ is the number of positive squaring basis vectors, $q$ is the number of negative squaring basis vectors, and $r$ is the number of null basis vectors. Hence \pyth{kingdon} supports all popular flavors of GA: \GR{p,q} aka VGA, \GR{p,q,1} aka PGA, and \GR{p+1,q+1} aka CGA.
        \item[Agnostic to input types] So long as the input datatypes support addition, subtraction, multiplication, division and square root operations, they can be used as multivector coefficients in  \verb|kingdon|. This includes popular datatypes such as \verb|PyTorch| tensors \cite{pytorch}, \verb|numpy| arrays \cite{numpy}, or \verb|sympy| symbolic expressions \cite{sympy}, but also the builtin complex numbers and even other \kingdon{} algebras.
        Hence, \verb|kingdon| plays well with others.
        \item[Interactive visuals] To really get a feel for your algorithms, it helps to be able to rapidly create interactive visuals. To this end \kingdon{} provides \ganja{} enabled graphics within the jupyter \cite{jupyter} ecosystem.
        \item[Leverage sparseness] While GA is famed for its multivectors, one rarely needs to perform operations between \emph{full} multivectors. 
        For example, in 3DPGA points, lines, and planes are encoded using trivectors, bivectors and vectors respectively, and therefore need far fewer coefficients than a full multivector. The most one ever needs in typical geometric applications is an even or odd multivector.
        The \verb|kingdon| library is designed to always take advantage of such input sparsity.
        \item[JIT \& CSE] All the common unary and binary operators of GA are symbolically optimized and Common Subexpression Elimination (CSE) is performed, to create Just-In-Time (JIT) compiled code  which is (close to) computationally optimal.
        \item[Clean API] We strive to keep the API user-friendly and unsurprising.
    \end{description}

\section{Code Design}

All elements of a geometric algebra are multivectors.
Kingdon's implementation of a \pyth{MultiVector} is as a dictionary-like object, meaning that it maps keys to values and implements the familiar \verb|.keys()|, \verb|.values()| and \verb|.items()| methods. 
The keys of a multivector are the binary representations of the basis blades, while the values are the corresponding multivector coefficients (these can be floats, symbols, arrays, etc.). 
The basis blades are ordered first by grade, and then lexicographic. 
To illustrate, the keys for \GR{2,0,1} are shown in \cref{tab:keys}.
\begin{table}[!htb]
    \centering
    \begin{tabular}{|l|l|l|l|l|l|l|l|l|}
    \hline
        blades & 1 & $\e{0}$ & $\e{1}$ & $\e{2}$ & $\e{01}$ & $\e{02}$ & $\e{12}$ & $\e{012}$ \\ \hline
        keys & 000 & 001 & 010 & 100 & 011 & 101 & 110 & 111 \\ \hline
    \end{tabular}
    \caption{Basis blade order and corresponding keys (in binary) for the algebra \GR{2, 0, 1}, made with the command \pyth{Algebra(2, 0, 1)}.}
    \label{tab:keys}
\end{table}

\noindent
Unlike dictionaries however, \verb|kingdon|'s multivectors do not support indexing by key in order to retrieve the corresponding multivector coefficient.\footnote{
For direct access users can use the \pyth{.values()} or \pyth{.items()} methods. Moreover, \texttt{kingdon} overloads the getattr syntax: 
i.e. for a multivector $x$,  \pyth{x.e1} returns the coefficient of the basis blade $\e{1}$.
(Dimensionality is a property of the elements, not of the space. Thus, a \texttt{MultiVector}  always returns $0$ as the coefficient of blades that are not part of its keys, even when that blade is not even part of the algebra used to create this multivector.)
}
Instead, Python indexing syntax is passed on to the multivector coefficients unseen, which enables powerful slicing to be performed on e.g. multivectors over arrays, as we shall see in \cref{lst:boids}.

Kingdon does not implement a full type system to type multivectors, but instead uses a single integer based on the keys to distinguish different kinds of multivectors. 
Specifically, the \pyth{type_number} of a \pyth{MultiVector} is generated by a bijective function which takes the keys, i.e. the non-zero basis blades, as input, and outputs an integer.
In the future this may be upgraded to a full type system that also incorporates the values in the type definition, in order to achieve even further performance improvements.\footnote{It can be meaningful to also include the values in the type definition, e.g. to distinguish points from arbitrary pseudovectors in PGA.}

\begin{table}
    \centering
    \scriptsize{
    \begin{tabular}{lc|c|l}
Operation & Expression & Infix & Inline \\
\hline
Geometric product & $ab$ & \verb`a*b` & \verb`a.gp(b)` \\
Inner & $a \cdot b$ & \verb`a|b` & \verb`a.ip(b)` \\
Scalar product & $\langle a \cdot b \rangle_0$ &  & \verb`a.sp(b)` \\
Left-contraction & $a \rfloor b$ & & \verb`a.lc(b)` \\
Right-contraction & $a \lfloor b$ & & \verb`a.rc(b)` \\
Outer (Exterior) & $a \wedge b$ & \verb`a ^ b` & \verb`a.op(b)` \\
Regressive & $a \vee b$ & \verb`a & b` & \verb`a.rp(b)` \\
Conjugate (sandwich) \verb`a` by \verb`b` & $b a \widetilde{b}$ & \verb`b >> a` & \verb`b.sw(a)` \\
Project \verb`a` onto \verb`b` & $(a \cdot b) \widetilde{b}$ & \verb`a @ b` & \verb`a.proj(b)` \\
Commutator of \verb`a` and \verb`b` & $a \times b = \tfrac{1}{2} [a, b]$ & & \verb`a.cp(b)` \\
Anti-commutator of \verb`a` and \verb`b` & $\tfrac{1}{2} \{a, b\}$ & & \verb`a.acp(b)` \\
Sum of \verb`a` and \verb`b` & $a + b$ & \verb`a + b` & \verb`a.add(b)` \\
Difference of \verb`a` and \verb`b` & $a - b$ & \verb`a - b` & \verb`a.sub(b)` \\
Reverse of \verb`a` & $\widetilde{a}$ & \verb`~a` & \verb`a.reverse()` \\
Squared norm of \verb`a` & $a \widetilde{a}$ & & \verb`a.normsq()` \\
Norm of \verb`a` & $\sqrt{a \widetilde{a}}$ & & \verb`a.norm()` \\
Normalize \verb`a` & $a / \sqrt{a \widetilde{a}}$ & & \verb`a.normalized()` \\
Square root of \verb`a` & $\sqrt{a}$ & & \verb`a.sqrt()` \\
Dual of \verb`a` & $a*$ & & \verb`a.dual()` \\
Undual of \verb`a` & & & \verb`a.undual()` \\
Grade \verb`k` part of \verb`a` & $\langle a \rangle_k$ & & \verb`a.grade(k)`
    \end{tabular}
    }
    \caption{Common unary and binary GA operators and their \texttt{kingdon} equivalents.}
    \label{tab:operators}
\end{table}

\begin{figure}[htb]
    \centering
    \includegraphics[width=0.8\linewidth]{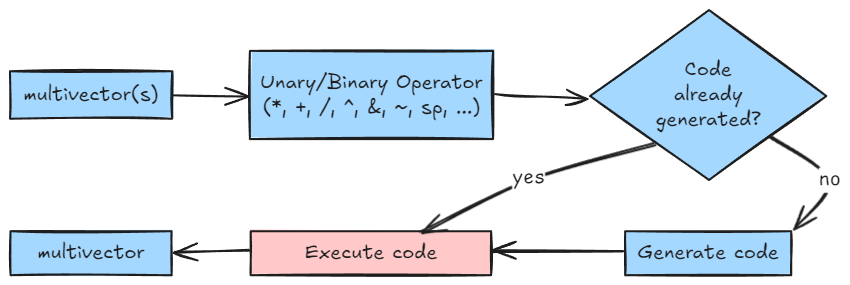}
    \caption{Diagramatic representation of the steps taken by \kingdon{} when performing operators between/on multivectors.}
    \label{fig:flowchart}
\end{figure}

\Cref{tab:operators} gives a list of common GA operators and their \kingdon{} equivalent, while \cref{fig:flowchart} outlines what happens when an operator is invoked.
First, the multivector delegates this call to the \pyth{Algebra} class, which checks if code for this operation with inputs of this type has already been generated. 
If not, the code is generated. 
Then the generated code is called with the \pyth{.values()} of the arguments, which returns the \pyth{.values()} of the output. Lastly the result is packaged into a new multivector.
The following example illustrates this in more detail.

\begin{example}\label{ex:codegen}
Consider how a bireflection $R$ in 2D rotates a line $x$ into a line $y$.
This is done using conjugation $y = R x \widetilde{R}$, which in \pyth{kingdon} is implemented using right-shift (\pyth{>>}).
We start by making a 2DVGA and the multivectors $R$ and $x$:
    \begin{python}
>>> from kingdon import Algebra
>>> alg = Algebra(2)
>>> R = alg.evenmv(name='R')
>>> x = alg.vector(name='x')
    \end{python}
In \cref{tab:operators} we see that the inline operator for conjugation is \pyth{sw}. An object by the same name exists on \pyth{Algebra} to generate and track all the generated code in its (currently empty) \pyth{operator_dict} attribute:
\begin{python}
>>> alg.sw.operator_dict
{}
\end{python}
Hence, no code is known after creation of a new algebra.
Both $R$ and $x$ have a \pyth{type_number} associated with them, which \kingdon{} uses to determine if it has seen a multivector of this kind before in this particular operation. 
At the time of writing the type numbers are $9$ and $6$ respectively.
We then apply the transformation $R$ to $x$ using the group action $R x \widetilde{R}$:
\begin{python}
>>> y = R >> x
>>> y
(R**2*x1 + 2*R*R12*x2 - R12**2*x1) e1 + (R**2*x2 - 2*R*R12*x1 - R12**2*x2) e2
\end{python}
Looking at \pyth{alg.sw.operator_dict} again, we find that it is no longer empty:
\begin{python}
{(9, 6): (6, <function codegen_sw_9_x_6(A, B)>)}
\end{python}
The entry says that the result of applying a type 9 to a type 6 multivector is again a 6, and contains the generated function to compute the result.
Closer inspection shows that the function \pyth{codegen_sw_9_x_6} is given by
\begin{python}[label=lst:codegen_sw,caption=Example of generated code.]
def codegen_sw_9_x_6(A, B):
    [a, a12] = A
    [b1, b2] = B
    x0 = a**2
    x1 = 2*a*a12
    x2 = a12**2
    return [b1*x0 - b1*x2 + b2*x1, -b1*x1 + b2*x0 - b2*x2]
\end{python}
Common Subexpression Elimination (CSE) has been applied to reduce code repetition, and \kingdon{} has taken advantage of the sparseness of the input and output, all in an effort to avoid unnecessary computations. 
\end{example}

Theoretically it might seem that generating code for every possible combination of basis elements will result in combinatorial hell, but in practice this is not an issue, as typical code will only involve a small number combinations.
It is a precondition however, for code generation to be sufficiently fast.
\Cref{tab:codegen} shows the execution times for the (geometric) product for numerical multivectors in the popular 2DPGA, 2DCGA, 3DPGA, and 3DCGA algebras.
It is important to note that the worst-case of a binary operation between two full multivectors $x$ and $y$ is atypical. 
The actual worst-case that one might encounter is e.g. the composition of two elements of the even subalgebra $R_1$ and $R_2$, and so this is listed as ``typical worst-case'' in the table. Lastly the table also includes a more typical scenario involving two vectors $u$ and $v$, and this is where \kingdon{} really starts to come into its own.\footnote{So as long as code generation happens in the blink of an eye, the resulting user experience will be snappy, compared to other libraries which spend a lot of time precompiling functionality the user did not ask for, and end up feeling sluggish.}
These results demonstrate that the initial code generation is sufficiently fast for users not to notice, and subsequent executions are 2-3 orders of magnitude faster.
If you expected the generated code to be faster, keep in mind that this is still pure Python, without further tricks to take the computations to a lower level, which we will address shortly.

\noindent
The high speeds during code generation are possible thanks to a rational polynomials class designed by Steven De Keninck for GAmphetamine.js \cite{gamphetamine}. 
A pure Python implementation thereof was found to be orders of magnitude faster than \text{SymPy} based code generation. 
Furthermore, early experiments found that by reimplementing the rational polynomials using cython at least one additional order of magnitude can be gained, but at the cost of adding cython as a dependency. This might be worth revisiting in the future.
\begin{table}[]
    \centering
    \begin{tabular}{l|>{\centering\arraybackslash}p{1.5cm}|>{\centering\arraybackslash}p{1.5cm}|>{\centering\arraybackslash}p{1.5cm}|>{\centering\arraybackslash}p{1.5cm}|>{\centering\arraybackslash}p{1.5cm}|>{\centering\arraybackslash}p{1.5cm}|}
    \cline{2-7}
        & \multicolumn{2}{c|}{typical} & \multicolumn{2}{c|}{typical worst-case} & \multicolumn{2}{c|}{absolute worst-case} \\
& \multicolumn{2}{c|}{\pyth{u * v}} & \multicolumn{2}{c|}{\pyth{R1 * R2}} & \multicolumn{2}{c|}{\pyth{x * y}} \\
\cline{2-7}
& $i = 1$ & $i > 1$ & $i = 1$ & $i > 1$ & $i = 1$ & $i > 1$ \\
\cline{2-7}
\GR{2, 0, 1} & 138 & 1.8 & 130 & 1.9 & 288 & 2.5 \\
\GR{3, 1} & 144 & 1.9 & 326 & 2.8 & 968 & 6.1 \\
\GR{3, 0, 1} & 143 & 1.9 & 294 & 2.5 & 813 & 5.1 \\
\GR{4, 1} & 189 & 2.1 & 974 & 6.2 & 3368 & 19 \\
\cline{2-7}
    \end{tabular}
    \caption{Timing (in microseconds) of code generation and first execution ($i = 1$) and subsequent executions ($i > 1$) of the product, for vectors $u, v$ (``typical''), even multivectors $R_1, R_2$ (``typical worst-case''), and full multivectors $x, y$ (``absolute worst-case''). Other products ($\wedge, \vee, \cdot$) involve smaller numbers of operations and are faster.}
    \label{tab:codegen}
\end{table}

\noindent
\Cref{ex:codegen} illustrated the first two techniques in \kingdon{}'s arsenal to generate efficient code: leverage sparseness and perform CSE. 
However, there is more to be done. 
After all, all the steps in \cref{fig:flowchart} take time, but the steps indicated in blue are essentially just glue, while the real computation only happens when executing the generated code.
There are two scenarios in which we clearly would like to avoid paying for the glue:
first, when doing the same operation many times for inputs of the same kind, e.g. rotating a point cloud; and second, when chaining together multiple operations, e.g. \pyth{(a | b) * b.inv()} to project $a$ onto $b$.
Kingdon has two tricks up its sleeve to address these common scenarios and to ensure high performance with minimal effort required from the user.
\paragraph{Broadcasting.}{
In order to perform the same operation many times, one should use a suitable array type, and let the broadcasting rules take care of the rest. For example, a pointcloud should not be realized as an array of multivectors,
\begin{python}
from random import uniform
alg = Algebra(2, 0, 1)
N = 1000
coords = [[1.0, uniform(-1, 1), uniform(-1, 1)] 
          for _ in range(N)]
pypoints = [alg.vector(coord).dual() for coord in coords]
\end{python}
but as a multivector over an array (numpy/PyTorch/etc.):
\begin{python}
npcoords = np.array(coords).T
nppoints = alg.vector(npcoords).dual()
\end{python}
Notice that we had to transpose such that \pyth{npcoords} has shape $(3, N)$ since \kingdon{} uses the first dimension for its unpacking (recall \cref{lst:codegen_sw}).
\Cref{tab:broadcast_timing} compares the performance of the two under rotation by $R = \exp(\e{12} / 2)$. 
It is overwhelmingly clear that a multivector over arrays is vastly superior to an array of multivectors, 
and unlike other Python GA packages, 
\kingdon{}'s design fully embraces this programming idiom, while simultaneously giving the users the freedom to use whichever array structure they prefer.
\begin{table}[h]
    \centering
    \begin{tabular}{|r|>{\centering\arraybackslash}p{1.5cm}|>{\centering\arraybackslash}p{1.5cm}|>{\centering\arraybackslash}p{1.5cm}|>{\centering\arraybackslash}p{1.5cm}|>{\centering\arraybackslash}p{1.5cm}|}
    \hline
    $N$                       & 1 & 10 & 100 & 1000 & 10000 \\
    \hline
    \texttt{R >> pypoints} & 8.3 & 36 & $ 2.1 \cdot 10^{2}$ & $ 2.1 \cdot 10^{3}$ & $ 2.2 \cdot 10^{4}$ \\
    \texttt{R >> nppoints} & 14 & 8.5 & 7.2 & 11 & 47 \\
    \hline
\end{tabular}
    \caption{Rotating an array of $N$ multivectors vs a multivector of length $N$ arrays. Timings (in microseconds) are the average of $1000$ runs. Timed in Python 3.10 on a MacBook Air M2.}
    \label{tab:broadcast_timing}
\end{table}
}
\paragraph{Register decorator.}{
It is common to string multiple operations together in order to form expressions. 
A simple example is PGA's projection formula $(a \cdot b) b^{-1}$: \pyth{project = lambda a, b:  (a | b) / b}.
In order to pay the glue only once, and to simply string together the generated code before the returning the result, we should define the function as
\begin{python}
@alg.register
def project(a, b):
    return (a | b) / b
\end{python}
Or we could make it even more performan
t by symbolically optimizing the expression with

\begin{python}
@alg.register(symbolic=True)
def project(a, b):
    return (a | b) / b
\end{python}
\begin{table}[H]
    \centering
    \begin{tabular}{|l|c|c|c|c|}
    \hline
        & \texttt{project} & \texttt{register(project)} & \texttt{register(symbolic=True)(project)} & \texttt{a @ b} \\
        \hline
        $i = 1$ & $ 1.9 \cdot 10^{3}$ & $ 2.0 \cdot 10^{3}$ & $ 4.1 \cdot 10^{3}$ & $ 1.0 \cdot 10^{3}$ \\
        $i > 1$ & $4.7$ & $3.2$ & $2.2$ & $1.9$ \\
        \hline
    \end{tabular}
    \caption{Timings (in microseconds) for the various projection implementations, for the projection of a single point onto a single line. In \kingdon{} projection is available as \pyth{a @ b}, which additionally assumes $b$ to be normalized for further performance improvement. Timings are the average over $10^5$ executions.}
    \label{tab:register}
\end{table}

\noindent
\Cref{tab:register} shows the times for the projection of a point onto a line in 2DPGA, and illustrates that significant performance improvements can be obtained with great ease.
However, symbolic optimization does incur extra costs upon code generation, so the user will have to consider this trade-off. But the \texttt{Algebra.register} decorator makes it easy for the user to experiment and reach an informed decision, with minimal effort on their part.
An interesting potential application where both broadcasting and pre-compiled expressions would come together is when implementing GA based neural networks, such as GCAN \cite{pmlr-v202-ruhe23a} or GATr \cite{NEURIPS2023_6f6dd92b}.
}

\paragraph{Remark.}{
One might expect that just-in-time libraries like numba could result in even more speed-ups. However, for large numpy arrays, almost all computational time is already spent in numpy code, and so numba optimization had no added value. Contrarily, for small amounts of data a speed-up due to numba can be observed, but at the cost of very expensive compilation times at first execution. And since small amounts of data are already processed in the blink of an eye even in pure python, numba has no added value.
An \pyth{Algebra} can be initiated with a \pyth{wrapper} function however, which will be applied as a decorator to all the generated functions, and so if desired numba can still be used as e.g. \pyth{Algebra(3, 0, 1, wrapper=numba.njit).}
}

\section{Kingdon Showcase}

This section highlights some examples of what is possible with \kingdon{}. Many more examples can be found and run directly online in the \href{https://tbuli.github.io/teahouse}{teahouse}.
\subsubsection{\kingdon{} embraces \ganja{}}

\begin{figure}[!htb]
    \centering
    \includegraphics[width=0.9\linewidth]{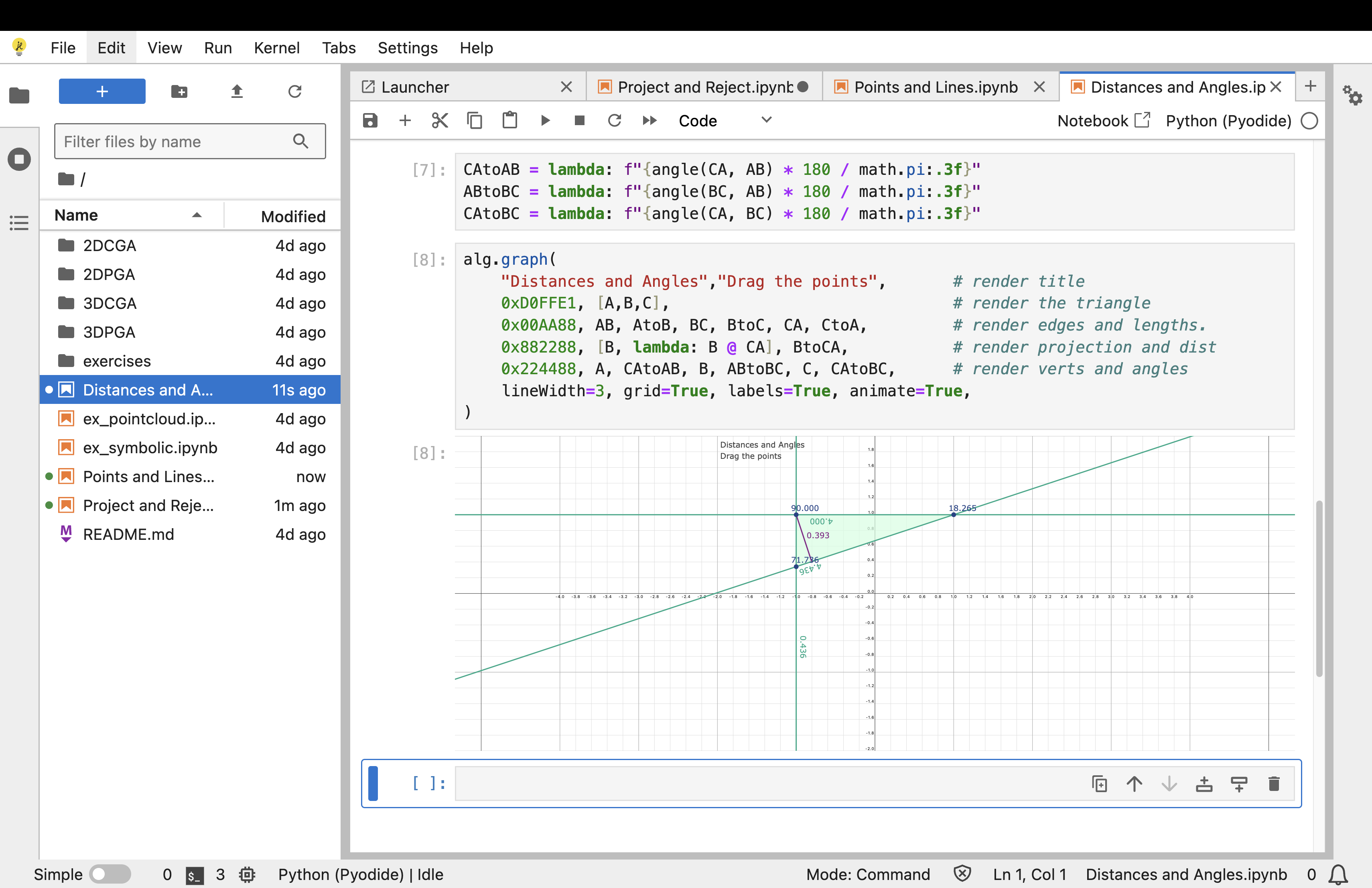}
    \caption{The \pyth{Algebra.graph} function creates a fully interactive ganja powered graph in jupyter notebooks. Moreover, the syntax is as identical to that of ganja as possible. Try it in the browser at \url{https://tbuli.github.io/teahouse}.}
    \label{fig:teahouse}
\end{figure}
Arguably the most popular GA package is \ganja{} by Steven De Keninck \cite{ganja}. It is not hard to see why: its \href{https://enkimute.github.io/ganja.js/examples/coffeeshop.html}{coffeeshop} makes it easy to play around with GA and instantly visualize the results. 
Ganja really puts the geometry into geometric algebra, which is something \kingdon{} aims to do for the Python ecosystem.

\noindent
However, ganja's graphs are not merely stills, but can be animated and are interactive. 
Porting these even more advanced ganja features to \kingdon{} requires a synchronized state of \kingdon{}'s Python multivectors with ganja's javascript multivectors.
This shared state was achieved by creating a widget using the excellent anywidget package \cite{manz2024anywidget,manz2024notebooks}.
The resulting \texttt{GraphWidget} allows users to have a fully interactive ganja experience from within jupyter notebooks, as shown in \cref{fig:teahouse}.
\\\\
\noindent
And just like \ganja{} has its \href{https://enkimute.github.io/ganja.js/examples/coffeeshop.html}{coffeeshop}, \kingdon{} has its \href{https://tbuli.github.io/teahouse}{teahouse}.

\subsubsection{Idiomatic \kingdon{} code}

Idiomatic \kingdon{} code leverages the package design to maximal effect. 
The core idea is to write high level code, capturing only the geometric relations, and to let the actual computation be handled by specialized packages like numpy or pytorch.
As an example of idiomatic \kingdon{} code, consider the implementation of boids algorithm, which can be found in the \href{{https://tbuli.github.io/teahouse/lab/index.html?path=2DPGA%2Fex_2dpga_boids.ipynb}}{teahouse}. The algorithm describes flocking behavior in e.g. birds, and consists of some simple rules which dictate that (1) boids strive to stay close to their neighbors, (2) align their direction with their neighbors, and (3) avoid collision.
First we need to setup the initial boid positions and velocities. 
Given the results of \cref{tab:broadcast_timing} we obviously want to do this using arrays.
The multivectors \texttt{boids} and \texttt{boidsvels} are pseudovectors of shape $(3, N_\text{boids})$, where $N_\text{boids}$ is the number of boids.
Now let us look at the actual \kingdon{} implementation of the first rule: boids should stay close to their neighbors. Thanks to the very hands-off approach of \kingdon{}, we can easily use the numpy masking syntax to select those boids which are in visibility range of a given boid, and make it fly towards the centre of mass of all its neighbours:
\begin{python}[caption={Make every boid fly towards the centre of mass of its neighbours. The example uses numpy, but would be identical, \emph{mutatis mutandis}, with e.g. pytorch.},label=lst:boids]
for i, boid in enumerate(boids):
    distance = (boid & boids).norm().e
    neighbors = distance < visual_range
    # Boids strive to stay together
    pos_avg = boids[neighbors].map(np.mean)
    boidvels[i] += (pos_avg - boid) * centering_factor
\end{python}
Notice how there is no mention of $x$ and $y$ coordinates, or the velocity components $v_x$ and $v_y$. Instead, the code is high level, as promised.
An example of the flocking behavior is shown in \cref{fig:boids}. For the full implementation of boids algorithm we refer to the \href{{https://tbuli.github.io/teahouse/lab/index.html?path=2DPGA%2Fex_2dpga_boids.ipynb}}{teahouse}.
\begin{figure}[!htb]
    \centering
    \includegraphics[width=\linewidth]{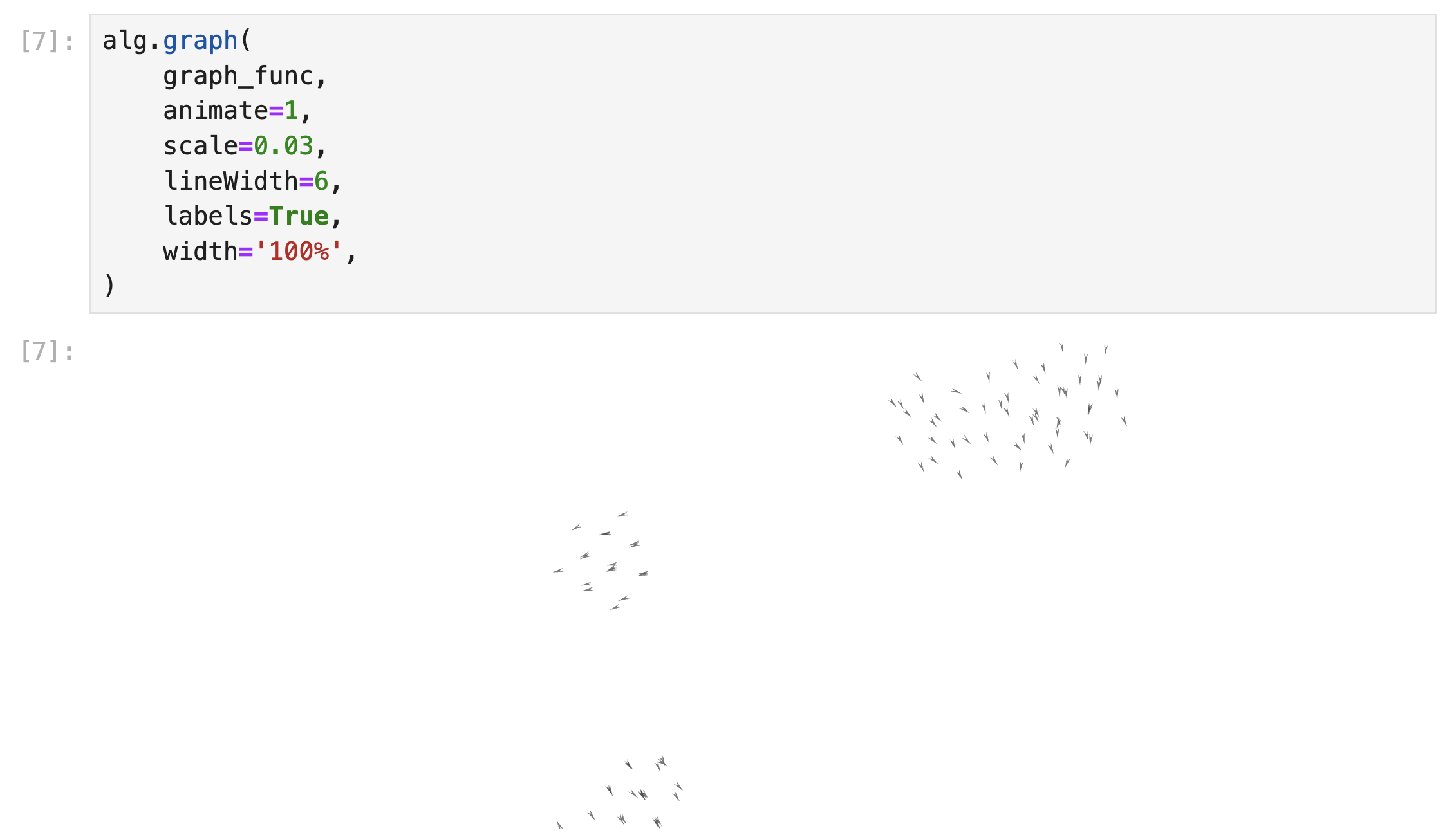}
    \caption{Screenshot from the boid example in the \href{{https://tbuli.github.io/teahouse}}{teahouse}.}
    \label{fig:boids}
\end{figure}

\subsubsection{Automatic Differentiation} \label{subsec_Aut_Dif}

Because \kingdon{} also supports algebras over algebras, it is straightforward to add automatic differentiation to existing code.
This was used to great effect in the AnCoOpt project at Flanders Make, which aims to optimize the torque and energy consumption of mechanisms (amongst other tasks, such as leveraging \kingdon{}'s computational efficiency to enable a Genetic Algorithm optimisation to be performed on these mechanisms).
Suppose we have a function \pyth{f(t, points)} 
that describes the evolution of the points in a mechanism as a function of time $t$, where $t$ is a scalar and points is a list of multivectors. 
In order to calculate the torque and energy we need to know the linear and angular velocities/accelerations of each bar in the mechanism \cite{mechanism}. 
How we compute these quantities in true PGA fashion without requiring a coordinate system for each bar will be the subject of an upcoming publication, but the first step is finding the first and second derivatives of each of the points w.r.t. time $t$.
Finding these derivatives is as easy as promoting the scalar time to a dual number:
\begin{python}
dualalg = Algebra(0, 0, 1)

taxis = np.linspace(0, 2 * np.pi)
t = dualalg.multivector(e=taxis, e0=1)
t = dualalg.multivector(e=t, e0=1)
new_points = f(t, points)
\end{python}
And that's it! We will now be able to extract the first and second derivatives for all the points at the time coordinates present in \texttt{taxis}, and with that for our entire mechanism. See \cref{fig:mechanism} for the first and second order derivatives of a pumping jack mechanism at a specific moment in time.

\begin{figure}
    \centering
    \includegraphics[width=0.5\linewidth]{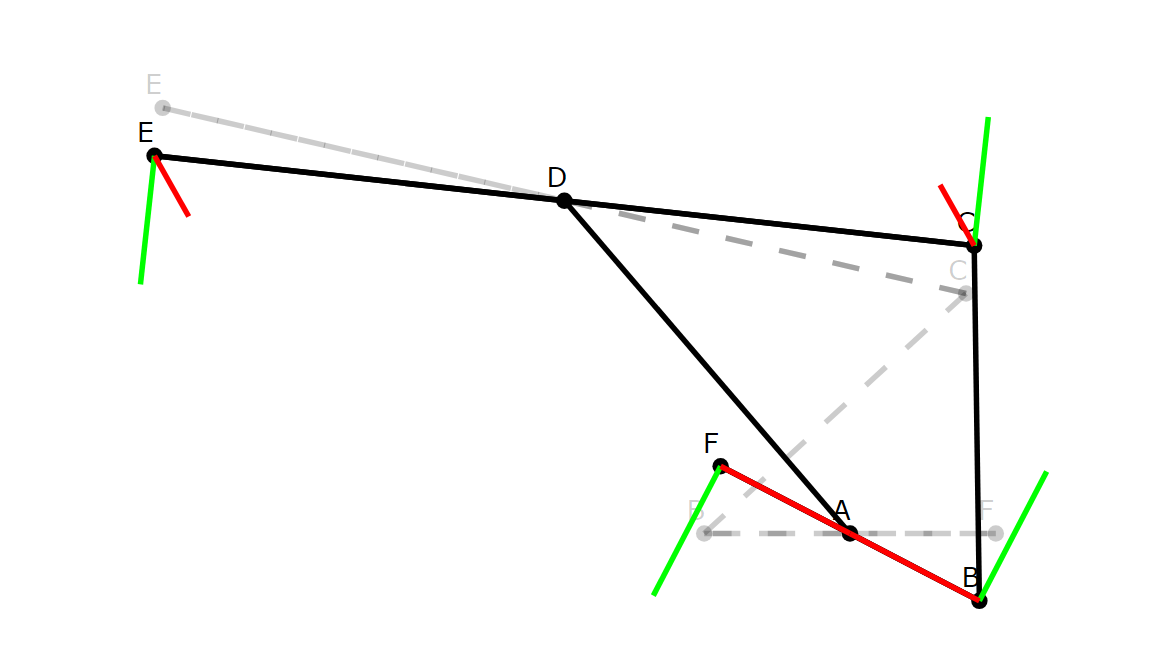}
    \caption{Pumping jack mechanism. Green: first derivative of a rod's end point. Red: second derivative of a rod's end point. The accelerations around the motor (in point A) balance each other out perfectly.}
    \label{fig:mechanism}
\end{figure}

\section{Conclusion}

Kingdon supports GA's of any signature, plays well with popular libraries such as numpy/sympy/PyTorch, and has unparalleled visualization capabilities owing to its \ganja{} integration.
Moreover, it allows the user to write high-level code that is nonetheless performant.

The symbiosis between \texttt{ganja.js}, \texttt{GAmphatamine.js}, and \kingdon{} has been remarkably successful in producing code that only captures mathematical statements, and hence looks similar across different programming languages. 
For example, code from the \ganja{} coffeeshop can be directly translated to \kingdon{} and vice versa, with the most important difference being the $\qquad\quad$ whitespace.
As Python and JavaScript are currently the two most popular programming languages, this covers a significant amount of the developers.

All of these features together make \kingdon{} an extremely well rounded GA package with the ambition to take GA in Python to a new level.

\subsubsection{\ackname}
Special thanks goes Steven De Keninck, the developer of \ganja{} and \texttt{GAmphetamine.js}.
Without our endless discussions about GA programming \kingdon{} would not be the package it is today.

The AnCoOpt project mentioned in subsection \ref{subsec_Aut_Dif} is a VLAIO TETRA project with Flanders Make as one of the project partners. Flanders Make is the strategic research centre for the Flemish manufacturing industry.

\bibliographystyle{splncs04}
\bibliography{biblio}

\end{document}

%% file: definitions.tex
\usepackage{physics}
\usepackage{mathtools}  

\newcommand{\GR}[2][{}]{\ensuremath{\mathbb{R}^{{#1}}_{#2}}}

\newcommand{\e}{\mathbf {e}_}            





%% file: main.bbl
\begin{thebibliography}{10}
\providecommand{\url}[1]{\texttt{#1}}
\providecommand{\urlprefix}{URL }
\providecommand{\doi}[1]{https://doi.org/#1}

\bibitem{pytorch}
Ansel, J., Yang, E., He, H., Gimelshein, N., Jain, A., Voznesensky, M., Bao, B., Bell, P., Berard, D., Burovski, E., Chauhan, G., Chourdia, A., Constable, W., Desmaison, A., DeVito, Z., Ellison, E., Feng, W., Gong, J., Gschwind, M., Hirsh, B., Huang, S., Kalambarkar, K., Kirsch, L., Lazos, M., Lezcano, M., Liang, Y., Liang, J., Lu, Y., Luk, C.K., Maher, B., Pan, Y., Puhrsch, C., Reso, M., Saroufim, M., Siraichi, M.Y., Suk, H., Zhang, S., Suo, M., Tillet, P., Zhao, X., Wang, E., Zhou, K., Zou, R., Wang, X., Mathews, A., Wen, W., Chanan, G., Wu, P., Chintala, S.: Pytorch 2: Faster machine learning through dynamic python bytecode transformation and graph compilation. In: Proceedings of the 29th ACM International Conference on Architectural Support for Programming Languages and Operating Systems, Volume 2. p. 929–947. ASPLOS '24, Association for Computing Machinery, New York, NY, USA (2024). \doi{10.1145/3620665.3640366}, \url{https://doi.org/10.1145/3620665.3640366}

\bibitem{NEURIPS2023_6f6dd92b}
Brehmer, J., de~Haan, P., Behrends, S., Cohen, T.S.: Geometric algebra transformer. In: Oh, A., Naumann, T., Globerson, A., Saenko, K., Hardt, M., Levine, S. (eds.) Advances in Neural Information Processing Systems. vol.~36, pp. 35472--35496. Curran Associates, Inc. (2023), \url{https://proceedings.neurips.cc/paper_files/paper/2023/file/6f6dd92b03ff9be7468a6104611c9187-Paper-Conference.pdf}

\bibitem{galgebra}
Bromborsky, A., Song, U., Wieser, E., Hadfield, H., {The Pygae Team}: pygae/galgebra (Jun 2020). \doi{10.5281/zenodo.3857096}, \url{https://doi.org/10.5281/zenodo.3857096}

\bibitem{gamphetamine}
De~Keninck, S.: Gamphetamine.js, \url{https://github.com/enkimute/GAmphetamine.js}

\bibitem{ganja}
De~Keninck, S.: ganja.js. Zenodo. https://doi.org/10.5281/ZENODO.3635774 (2020). \doi{10.5281/ZENODO.3635774}, \url{https://zenodo.org/record/3635774}

\bibitem{jupyter}
Granger, B.E., Pérez, F.: Jupyter: Thinking and storytelling with code and data. Computing in Science \& Engineering  \textbf{23}(2),  7--14 (2021). \doi{10.1109/MCSE.2021.3059263}

\bibitem{python_clifford}
Hadfield, H., Wieser, E., Arsenovic, A., Kern, R., {The Pygae Team}: pygae/clifford. \doi{10.5281/zenodo.1453978}, \url{https://doi.org/10.5281/zenodo.1453978}

\bibitem{numpy}
Harris, C.R., Millman, K.J., van~der Walt, S.J., Gommers, R., Virtanen, P., Cournapeau, D., Wieser, E., Taylor, J., Berg, S., Smith, N.J., Kern, R., Picus, M., Hoyer, S., van Kerkwijk, M.H., Brett, M., Haldane, A., del R{\'{i}}o, J.F., Wiebe, M., Peterson, P., G{\'{e}}rard-Marchant, P., Sheppard, K., Reddy, T., Weckesser, W., Abbasi, H., Gohlke, C., Oliphant, T.E.: Array programming with {NumPy}. Nature  \textbf{585}(7825),  357--362 (Sep 2020). \doi{10.1038/s41586-020-2649-2}, \url{https://doi.org/10.1038/s41586-020-2649-2}

\bibitem{manz2024anywidget}
Manz, T., Abdennur, N., Gehlenborg, N.: anywidget: reusable widgets for interactive analysis and visualization in computational notebooks. Journal of Open Source Software  \textbf{9}(102), ~6939 (2024). \doi{10.21105/joss.06939}, \url{https://doi.org/10.21105/joss.06939}, publisher: The Open Journal

\bibitem{manz2024notebooks}
Manz, T., Gehlenborg, N., Abdennur, N.: Any notebook served: authoring and sharing reusable interactive widgets. In: Proceedings of the 23rd {Python} in {Science} {Conference} (Jul 2024). \doi{10.25080/NRPV2311}, \url{https://doi.curvenote.com/10.25080/NRPV2311}

\bibitem{sympy}
Meurer, A., Smith, C.P., Paprocki, M., \v{C}ert\'{i}k, O., Kirpichev, S.B., Rocklin, M., Kumar, A., Ivanov, S., Moore, J.K., Singh, S., Rathnayake, T., Vig, S., Granger, B.E., Muller, R.P., Bonazzi, F., Gupta, H., Vats, S., Johansson, F., Pedregosa, F., Curry, M.J., Terrel, A.R., Rou\v{c}ka, v., Saboo, A., Fernando, I., Kulal, S., Cimrman, R., Scopatz, A.: Sympy: symbolic computing in python. PeerJ Computer Science  \textbf{3}, ~e103 (Jan 2017). \doi{10.7717/peerj-cs.103}, \url{https://doi.org/10.7717/peerj-cs.103}

\bibitem{mechanism}
Oosterwyck, N.V., Vanbecelaere, F., Knaepkens, F., Monte, M., Stockman, K., Cuyt, A., Derammelaere, S.: Energy optimal point-to-point motion profile optimization. Mechanics Based Design of Structures and Machines  \textbf{52}(1),  239--256 (2024). \doi{10.1080/15397734.2022.2106241}, \url{https://doi.org/10.1080/15397734.2022.2106241}

\bibitem{pmlr-v202-ruhe23a}
Ruhe, D., Gupta, J.K., De~Keninck, S., Welling, M., Brandstetter, J.: Geometric clifford algebra networks. In: Krause, A., Brunskill, E., Cho, K., Engelhardt, B., Sabato, S., Scarlett, J. (eds.) Proceedings of the 40th International Conference on Machine Learning. Proceedings of Machine Learning Research, vol.~202, pp. 29306--29337. PMLR (23--29 Jul 2023), \url{https://proceedings.mlr.press/v202/ruhe23a.html}

\end{thebibliography}
